\documentclass[twocolumn]{aa}
\usepackage{graphicx}
\usepackage{txfonts}
\def\figdir{./figs}
\def\hfigsize{4.0cm}
\def\figsize{8.0cm}
\def\etal{et~al.}
\begin{document}
   \title{The role of the time step
          and overshooting in the modelling of PMS evolution: the case of 
EK Cephei}
%   On the role of the integration time step for the calibration of 
%          the young binary EK Cephei}
   \titlerunning{Time step and overshooting during the PMS}

   \author{J. P. Marques\inst{1,2}
          \and
           J. Fernandes\inst{2,3}
	  \and
	   M. J. P. F. G. Monteiro\inst{1,4}
          }
   \authorrunning{J. P. Marques \etal}

   \offprints{}

   \institute{Centro de Astrof\'\i sica da Universidade do Porto, 
	      Rua das Estrelas, 4150-762 Porto, Portugal
         \and
             Grupo de Astrof\'\i sica da Universidade de Coimbra, Observat\'orio 
	Astron\'omico da Universidade de Coimbra, Santa Clara, Coimbra, Portugal	 
	 \and
	     Departamento de Matem\'atica da FCTUC, Coimbra, Portugal
	 \and
	     Departamento de Matem\'atica Aplicada, Faculdade de Ci\^encias
             da Universidade do Porto, Portugal
	     }

   \date{Received ; accepted }

   \abstract{
\object{EK~Cephei} (HD~206821) is a unique candidate to test predictions based
on stellar evolutionary models. It is a double-lined detached eclipsing binary 
system with accurate absolute dimensions available and a precise determination
of the metallicity. Most importantly for our work, its low mass (1.12
$M_{\sun}$) component appears to be in the pre-main sequence (PMS) phase.

We have
produced detailed evolutionary models of the binary \object{EK~Cep} using the
CESAM stellar evolution code (Morel \cite{morel97}). A $\chi^2$-minimisation was
performed to derive the most reliable set of modelling parameters (age, 
$\alpha_{\rm A}$, $\alpha_{\rm B}$ and $Y_{\rm i}$). We have found that an 
evolutionary age of about 26.8 Myrs fits both components in the same isochrone.
The positions of EK~Cep~A and B in the HR diagram are consistent (within the
observational uncertainties) with our results.

Our revised
calibration shows clearly that \object{EK~Cep~A} is in 
the beginning of the main 
sequence, while \object{EK~Cep~B} is indeed a PMS star. 
Such a combination allows for a 
precise age determination of the binary, and provides a strict test of the 
modelling. In particular we have found that the definition of the time step in 
calculating the PMS evolution is crucial to reproduce the observations. A 
discussion of the optimal time step for calculating PMS evolution is 
presented.

The fitting to the radii of both components is a more difficult task; although 
we managed to do it for EK~Cep~B, EK~Cep~A has a lower radius than our 
best models.

We further studied the effect of the inclusion of a
moderate convective overshooting; the calibration of 
the binary is not significantly altered, but the effect of the inclusion of
overshooting can be dramatic in the approach to the
main sequence of stars with masses high enough to burn hydrogen
through the CNO cycle on the main sequence.

}

   \maketitle
%
%______________________________________________________________________
\section{Introduction}

It is very well known that the interior structure of an evolved star is 
fixed by the initial mass $M_{\star}$, the initial chemical composition (initial 
abundances in hydrogen, helium and metals, $X_{\rm i}$, $Y_{\rm i}$ and $Z_{\rm 
i}$,
respectively) and the age $t_{\star}$. So, except for the Sun, modelling a 
single star using stellar evolutionary models on the HR diagram is not a 
closed problem because the number of parameters to be determined is larger than 
the observational constraints, the present observed luminosity and effective 
temperature.

This difficulty is increased due to uncertainties on the 
physics of the models used to describe the stellar interior/external 
structure. Currently these uncertainties are modelled by free parameters 
such as the 
diffusive and mass loss coefficient or the mixing length 
and overshooting parameters for the convection in the MLT approximation 
(Mixing Length Theory).

Some stellar multiple systems are in excellent position to avoid these 
difficulties. Good determinations of stellar masses are possible for 
the components of a given
binary. Based on the reasonable assumption of a common origin for both
components (which yields the same initial chemical composition and age), the
problem of modelling both stars of a binary system is reduced to the 
determination of three
so-called stellar modelling parameters, namely, $Y_{\rm i}$, $Z_{\rm i}$,
$t_{\star}$ (since $X{+}Y{+}Z{=}1$ we do not need to determine $X_{\rm i}$ 
independently). Moreover, if the stars are 
young enough (so that diffusion processes have not had time to change the 
surface composition significantly) and the present surface metallicity of the 
stars is also known (and is, therefore, equal to the initial metallicity), then 
$(Z/X)_{\rm i}$ is known. So, the number of observables (five: the 
effective temperatures and luminosities for both stars plus the metallicity of 
the system) is 
higher than the number of modelling parameters. This allows us to include 
unknown parameters related to the description of the interior physics. For low 
mass stars, without mass-loss or strong rotation, the convection is the 
main source of uncertainties in the description of the stellar interior. So we 
choose to include the mixing length parameter for each component of a binary 
system $\alpha_{\rm 
A}$ and $\alpha_{\rm B}$. A number of binary systems have been 
calibrated in this way, e.g. \object{$\alpha$~Cen} (Noels \etal\  
\cite{noels91}; Morel \etal\  \cite{morel00b}) and \object{$\iota$~Peg} (Morel 
\etal\  \cite{morel00a}); all these stars are, however, main sequence (MS) or 
post-main sequence stars.

A good candidate binary system for calibration must have 
well-determined luminosities, effective temperatures, metallicities and 
dynamical masses. Pre-main sequence (PMS) binaries that have all the required 
characteristics are very rare at present (see Palla \& Stahler \cite{palla01},
for some examples of PMS binaries, and Lee \etal\  \cite{lee94},
for lithium abundances for some of them).

%===============
\begin{table}
\caption{Properties of \object{EK~Cep} (HD~206821).
All data except chemical composition are from Andersen (\cite{andersen91}) and
Popper (\cite{popper87}); chemical compositions are from Mart\'{\i}n \& Rebolo
(\cite{martin93}). See also Tomkin (\cite{tomkin83}), Ebbighausen
(\cite{ebbighausen66}), and Hill \& Ebbighausen (\cite{hill84}).}\label{tab:obs}
\begin{center}
\begin{tabular}{lcc}
\hline
\noalign{\smallskip}
 & EK~Cep~A & EK~Cep~B \cr
\noalign{\smallskip}
\hline
\noalign{\smallskip}
$M_{\star}/M_{\sun}$ & 2.029$\pm$0.023 & 1.124$\pm$0.012 \cr
$\log T_{{\rm eff},\star}$ (K) & 3.954$\pm$0.010 & 3.756$\pm$0.015 \cr
$\log \left( L_{\star}/L_{\sun} \right)$ & 1.17$\pm$0.04 & 0.21$\pm$0.06 \cr
$R_{\star}/R_{\sun}$ & 1.579$\pm$0.007 & 1.315$\pm$0.006 \cr
$\log g$ (cgs) & 4.349$\pm$0.010 & 4.251$\pm$0.006 \cr
Sp. Type & A1.5V & G5Vp \cr
[Fe/H] & ... & +0.07$\pm$0.05 \cr
\noalign{\smallskip}
\hline
\end{tabular}
\end{center}
\end{table}
%===============

The double-lined eclipsing binary system \object{EK~Cep} (HD~206821) has
these unique characteristics. Accurate absolute dimensions are available
(Andersen \cite{andersen91}; Popper \cite{popper87}) and a determination of 
the surface metallicity was made (Mart\'\i n \& Rebolo \cite{martin93}). Most 
importantly, the radii of the two stars are much closer than expected given 
their mass ratio. This indicates that EK~Cep~B is still contracting towards the 
main-sequence. At the same time, the higher mass of the primary implies that the 
star must be already in its main sequence evolution.
All these make \object{EK~Cep} a perfect candidate to test
theoretical models of pre-ZAMS solar-type stars.
A summary of the observational characteristics of this binary
is given in Table~\ref{tab:obs}.
 
Several authors have tried to model this system. A problem frequently found 
is 
that EK~Cep~A has a smaller radius (or that 
EK~Cep~B has a bigger radius) than that predicted by the models, making it 
difficult to fit both components to the observations in the same isochrone.
Claret \etal\ (\cite{claret95}) have 
computed models that agree reasonably well with the observations; however, 
the radius of EK~Cep~B is somewhat higher than their models predict
(a problem also 
found in the models used for comparison with the observations in Mart\'\i n \& 
Rebolo \cite{martin93}). Y\i ld\i z (\cite{yildiz03}) computed models using a
higher metallicity than Claret \etal\ 
 and a fast rotating  core for EK~Cep~A in order to 
lower its luminosity and radius; this way, models that fit the observed radii 
and luminosities of the EK~Cep system can be computed. The metallicity used in 
Y\i ld\i z (\cite{yildiz03}; $Z\sim 0.04$) is much higher, however, 
than that given in 
Mart\'\i n \& Rebolo (\cite{martin93}; $Z \sim 0.02$). The values for the iron 
abundance given by these authors spread somewhat, making a very precise 
determination 
 of this abundance difficult; nothing in these values comes close, however, 
to the value of metallicity used in Y\i ld\i z (\cite{yildiz03}).

Our atempt to model this system focuses, instead, on the physical input that 
might affect the PMS evolution. Our own PMS models, computed using the CESAM 
stellar evolution code (Morel \cite{morel97}), suggested that there are effects 
in the PMS evolution due to the physical inputs (overshooting, time step) that 
have been ignored or too quickly dismissed in the previous works.

%______________________________________________________________________
\section{Physical ingredients}

Models were computed using the CESAM stellar evolution code (Morel 
\cite{morel97}). Typically, each model is described by about 600 shells, and an 
evolution by about 400-600 models. We restricted the maximum time step of the 
evolution: for EK~Cep~A, the maximum time step was kept at about 0.017 Myrs, 
while for EK~Cep~B we allowed the time step to reach 0.075 Myrs.
We justify these values for the time step in section~\ref{sec:time}.

Each evolution is initialised with an homogeneous, fully convective 
model in quasi-static contraction (Iben \cite{iben65}), with a central 
temperature inferior to the ignition temperature of the deuterium. We shall call 
the ``age'' of the model the time elapsed since initialisation. This initial 
model for PMS evolution is not a realistic assumption, since stars do not form 
by the homologous contraction of the pre-stellar cloud. Instead, a hydrostatic 
core first forms, which accretes mass from the parental cloud until either 
stellar winds disperse the cloud or the cloud material exhausts. Then the star 
becomes optically visible for the first time, since no more surrounding material 
obscures it: the star is born (see Stahler \cite{stahler83}; Palla \& Stahler 
\cite{palla91}, \cite{palla92}). If we wanted to
model very young stars, the problem of the initial conditions would be very
important; for \object{EK~Cep}, however, the initial conditions are 
irrelevant since it is much older than 1 Myr. Our ``age'' of the model, although 
not a real age of the star, is very close to it.

The model of the zero-age main-sequence (ZAMS) is defined as the first model 
where nuclear reactions account for more than 99\% of the energy generation.

We used the OPAL equation of state (Rogers \etal\  \cite{rogers96};
we used the 2001 version of the OPAL
tables which goes down to 2000 K) and the opacities of
Iglesias \& Rogers (\cite{iglesias96}),
complemented, at low temperatures,
by the Alexander \& Fergusson (\cite{alexander94}) opacities.

The temperature gradient in convection zones is computed using the standard
mixing-length theory (B\"ohm-Vitense \cite{vitense58}).
The mixing length is defined as $l {=} \alpha_\star H_P$,
$H_P$ being the local pressure scale height,
$H_P {=} -{\rm d} r /{\rm d}\ln P$.

The nuclear network we use contains the following species:
\element[][1]{H}, \element[][2]{H}, \element[][3]{He}, \element[][4]{He},
\element[][7]{Li}, \element[][7]{Be}, \element[][12]{C}, \element[][13]{C},
\element[][14]{N}, \element[][15]{N}, \element[][16]{O}, \element[][17]{O},
\element[][9]{Be}, and an ``extra'' fictitious non-CNO heavy element which
complements the mixture; this element has atomic mass 28 and charge 13.
Deuterium and lithium burning are taken into account, as well as the
most important reactions of the PP+CNO cycles. The nuclear reaction rates are
taken from the NACRE compilation (Angulo \etal\  \cite{angulo99}).

Given the expected young age of the binary, we do not consider diffusion
processes.
However, these effects must be taken into account if we would follow
the evolution of the surface chemical composition.
If the surface chemical composition were known with high accuracy
for both components, the binary would be an excellent candidate to test
diffusion in the models.

%______________________________________________________________________
\section{The fitting method}\label{sec:cal}

The calibration of a binary system consists of adjusting the stellar modelling
parameters so that the models reproduce the observational data of the stars. 
The effective temperature and luminosity of a model, for a given mass and
fixed input physics (EOS, opacity, nuclear energy generation rates),
depend on the
modelling parameters: the age $t_{\star}$, initial helium abundance $Y_{\rm i}$
and mixing length parameter $\alpha_{\star}$.

%===============
\begin{figure}[ht]
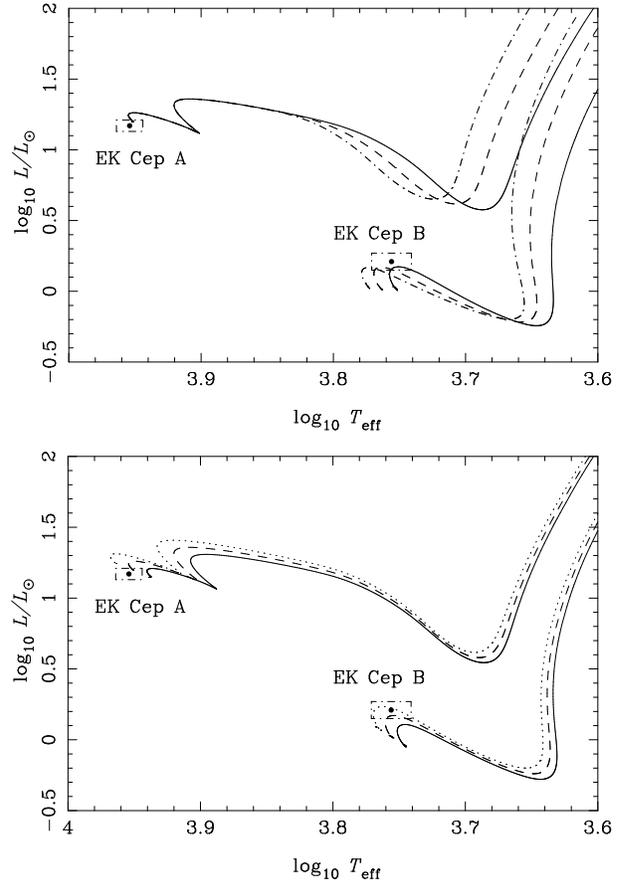

\centering
\begin{tabular}{c}
\includegraphics[height=\figsize,angle=-90]{\figdir/0693fg1a.ps} \\
\includegraphics[height=\figsize,angle=-90]{\figdir/0693fg1b.ps} 
\end{tabular}
\caption{Upper panel: variation of the evolutionary tracks with the mixing 
length parameter $\alpha$. Full lines are evolutionary tracks for 
$\alpha{=}1.3$; dashed lines are for $\alpha{=}1.6$; and dotted lines are
for $\alpha{=}1.9$
Lower panel: variation of the evolutionary tracks with the initial
helium abundance. Full lines are evolutionary tracks for $Y{=}0.24$; 
dashed lines are for $Y{=}0.26$; and dotted lines are for $Y{=}0.28$.
}\label{fig:var}
\end{figure}
%===============

The effect that the variation of the initial helium abundance $Y_{\rm i}$ and 
mixing length parameter $\alpha_{\star}$ has on the evolutionary tracks can be
seen in Fig.~\ref{fig:var}. An increase in $\alpha_{\star}$ moves the
evolutionary tracks in the HR diagram towards larger effective temperatures at
approximately constant luminosity, for stars with convective envelopes.
An increase of the 
efficiency of the convection deepens the base of the convective envelope, while 
it does not affect the core. The temperature of the base of the external 
convective zone therefore increases, and the adjustment of the adiabat to a 
higher temperature at its base causes the effective temperature to increase, 
leaving the luminosity almost unchanged.

Fig.~\ref{fig:zc} shows the evolution of the convective zones of EK~Cep~A and B.
The convective envelope of EK~Cep~A disappears as the star approaches the main
sequence, hence the small effect of the variation of $\alpha_{\star}$ in the 
tracks of EK~Cep~A on the later phases of PMS evolution. 

A larger helium abundance shifts the evolutionary tracks on the HR diagram 
towards higher effective temperature and higher luminosity (see lower 
panel of Fig.~\ref{fig:var}). That is because a larger helium abundance causes 
the mean molecular weight to increase, which decreases the pressure. To maintain
quasi-static equilibrium, the temperature must increase, which causes an
increase in the energy generation rate and, thus, an increase in luminosity.
This increase is more obvious in main sequence stars because of
the large power-law dependence of the nuclear reaction rates on temperature. On
the other hand, the gravitational energy generation rate increases only linearly 
with temperature and the rise in luminosity due to the larger helium abundance 
is smaller for PMS stars.

We can, then, produce a stellar model with a higher effective temperature by
increasing the parameter $\alpha_\star$;
to change only the luminosity of the model,
we can change simultaneously $Y_{\mathrm{i}}$ and $\alpha_\star$.
To obtain observables as close as possible to the observations,
we use the $\chi^2$ fitting developed by
Lastennet \etal\  (\cite{lastennet99}; see also Morel \etal\  \cite{morel00b}).
It corresponds to minimising the following functional:
\begin{equation}
\chi^2 {=} \! \sum_{{\star}=A,B}\left\{\left[
   \frac{\log T_{{\rm eff},\star}^{\rm mod} - 
         \log T_{{\rm eff},\star}}
   {\sigma(\log T_{{\rm eff},\star})}
\right]^2 \!\! {+} \left[
   \frac{\log L_{\star}^{\rm mod} - \log L_{\star}}
   {\sigma(\log L_{\star})}
\right]^2\right\} \! .
\label{eq:chi2}
\end{equation}
This functional depends on the modelling parameters through
$\log T_{{\rm eff},{\star}}^{\rm mod}$ and $\log L_{\star}^{\rm mod}$.
For a grid of modelling
parameters we have computed the evolution of models with the masses and
metallicities of EK~Cep~A and B. The functional $\chi^2$ was computed for
each of these models using the above equation. The parameters that minimised 
$\chi^2$ were selected as the solution.

%______________________________________________________________________
\section{Results}

%===============
\begin{table}[ht]\caption{Modelling parameters for a model of EK~Cep 
lying within the uncertainty boxes for $\log T_{\rm eff,\star}$ and $\log 
L_{\star}/L_{\sun} $ (see text). The parameters for the Sun are for a
calibrated solar model using the same input physics. We do not present the 
value for $\alpha_{\rm A}$ because there is little dependence of the
location of the model of EK~Cep~A on this parameter
(see Fig.~\ref{fig:var}).}\label{tab:calib}
\centering
\begin{tabular}{lcc}
\hline
\noalign{\smallskip}
 & EK~Cep & Sun \cr
\noalign{\smallskip}
\hline
\noalign{\bigskip}
%& & \cr
$\alpha_{\rm B}$ & $1.363^{+0.307}_{-0.173}$ & 1.632 \cr \noalign{\bigskip}
$Y_{\rm i}$ & $0.261^{+0.005}_{-0.006}$ & 0.265 \cr \noalign{\bigskip}
$t_{\star}$ & $26.8^{+1.6}_{-1.9}$ Myrs & 4.52 Gyrs \cr \noalign{\medskip}
\noalign{\smallskip}
\hline
\end{tabular}
\end{table}
%===============

%===============
\begin{figure}[ht]
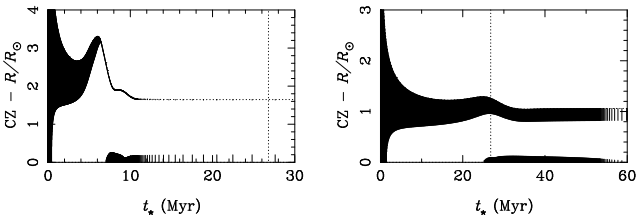

\centering
\begin{tabular}{lr}
\includegraphics[height=\hfigsize,angle=-90]{\figdir/0693fg2a.ps} &
\includegraphics[height=\hfigsize,angle=-90]{\figdir/0693fg2b.ps} \\
\end{tabular}
\caption{Convective zones and radii of EK~Cep~A (left panel) and B (right
panel). Vertical
lines represent the extent of convective zones; the vertical dotted line
in both panels represents the best model, with an age of 26.8~Myrs. 
}\label{fig:zc}
\end{figure}
%===============

The modelling parameters that minimise $\chi^2$ are given in
Table~\ref{tab:calib},
together with the same parameters for a calibrated solar model
using the same input physics.
The uncertainties are derived by changing one parameter at a time around the
calibrated value, keeping all the others constant. The value for
$\alpha_{\rm A}$ is not listed in Table~\ref{tab:calib}, since the dependence of 
the evolutionary tracks on this parameter is very weak (see 
section~\ref{sec:cal}, and in particular Fig.~\ref{fig:var}).
%This is also the reason why we do not present%the value of this parameter in 
%Table~\ref{tab:calib}. 
We used $\alpha_{\rm A}=1.340$ to produce these models.

Our calibrated value of $\alpha_{\rm B}$ is lower than the value of
$\alpha_{\star}$ for calibrated models of solar-mass stars (see, for example,
Morel \etal\  \cite{morel00b} for the case of \object{$\alpha$~Cen}, as well as
the value for the solar case we present in Table~\ref{tab:calib}). Typically,
one has $\alpha_{\star}\sim$~1.6~--~1.8 for calibrated solar-type stars.
D'Antona \& Montalb\'an (\cite{dantona03}) argue that in order to reproduce
both the lithium depletion and the location of PMS tracks
in the HR diagram, convection must be highly inefficient during the PMS;
that is, it must have a very low $\alpha_{\star}\sim 1$. Our
value of $\alpha_{\rm B}$ is between that appropriate for a calibrated MS model
and the value argued by D'Antona \& Montalb\'an for PMS models when they are
depleting their lithium. This could reflect the fact that \object{EK~Cep~B} is
in an evolutionary phase between the main lithium depletion phase (which occurs
for stars with this mass around the time a radiative core starts to
develop) and the ZAMS.

%===============
\begin{figure}[ht]
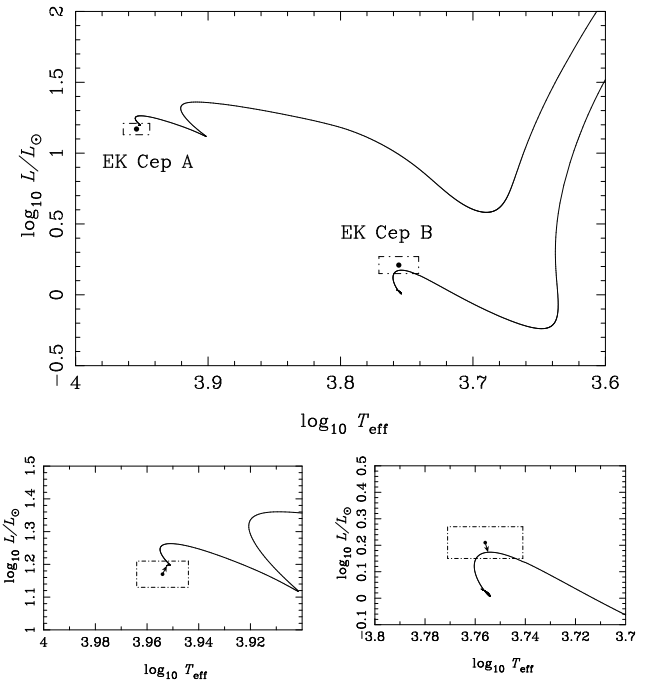

   \centering
\begin{tabular}{lr}
   \multicolumn{2}{c}
   {\includegraphics[height=\figsize,angle=-90]{\figdir/0693fg3a.ps}} \\
   \includegraphics[height=\hfigsize,angle=-90]{\figdir/0693fg3b.ps} &
   \includegraphics[height=\hfigsize,angle=-90]{\figdir/0693fg3c.ps} \\
\end{tabular}
\caption{Evolutionary tracks of EK~Cep~A and B on the HR diagram. The
uncertainty boxes of both stars are also shown.
%and the crosses on the tracks indicate the age in Myrs. 
In the lower left panel, an enlargement of the locus
of EK~Cep~A is shown; the arrow shows the location on the HR diagram of the
calibrated model.}\label{fig:hr}
\end{figure}
%===============

The sequences of models for the evolution determined with
the best set of parameters are shown in Fig.~\ref{fig:zc} and Fig.~\ref{fig:hr}.
Both components of EK~Cep have a convective core;
EK~Cep~A is a main sequence star with an energy production
mostly due to the CNO cycle,
while EK~Cep~B has just acquired a convective core due to the
start of \element[][12]{C} burning.

This convective core in EK~Cep~B will be lost shortly after the arrival
on the ZAMS. The ZAMS occurs at an evolutionary age of
$t_{\rm zams,A}{=}10.8$~Myrs and $t_{\rm zams,B}{=}51.3$~Myrs
for EK~Cep~A and B, respectively.

From our estimated age of around 27~Myrs, follows that EK~Cep~A
is already in the main-sequence, while EK~Cep~B is indeed a PMS
star (see Figs~\ref{fig:zc} and \ref{fig:hr}).
The fact that both components are in different stages of evolution
allows for a precise age determination, as the solution minimising
Eq.~\ref{eq:chi2} is strongly dependent on $t_\star$.

%______________________________________________________________________
\section{The problem of the radius}\label{sec:radius}

Our best models for EK~Cep~A and B do not fit, however, the observed radii. 
Since EK~Cep is an eclipsing binary, the radii of both stars are known with 
great precision; the fact that these models do not fit the radii is, therefore, 
a serious problem.

%===============
\begin{figure}[ht]
   \centering
   \includegraphics[height=\figsize,angle=-90]{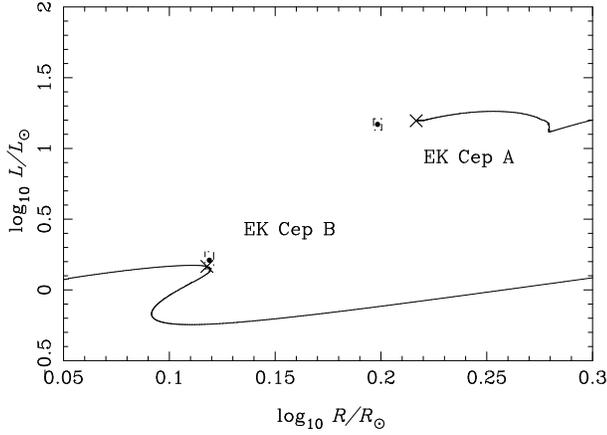}
\caption{Evolutionary tracks of EK~Cep~A and B on the $\log R/R_{\sun}$--$\log 
L/L_{\sun}$ plane. These tracks were calculated using 
$\alpha_{\rm B}=1.28$. Crosses show the location of models with 
26~Myrs.}\label{fig:lr}
\end{figure}
%===============

We can fit the radius of EK~Cep~B to the observations using a lower $\alpha_{\rm 
B}$ in the models. Fig.~\ref{fig:lr} shows the evolutionary tracks for EK~Cep~A 
and B on the $\log R/R_{\sun}$--$\log L/L_{\sun}$ plane, using $\alpha_B=1.28$ 
(all other parameters being the same as in the previous section). However, the 
same cannot be done for EK~Cep~A since the model depends weakly on the 
mixing length parameter $\alpha$.

We performed the fitting decribed in section \ref{sec:cal} using the effective
temperature and radius as the observables. The parameters that minimise 
$\chi^2$ are now
$Y_{\rm i}=0.331$, $t_{\star}=13.9$~Myrs and $\alpha_{\rm B}=1.68$. 
The helium abundance seems very large for the metallicity of EK~Cep, but it 
reproduces the results presented in Lastennet \etal\ (\cite{lastennet03}) for 
the binary \object{UV Psc}. However, a
model of EK~Cep~A constructed using these parameters still has 
too large a radius.

%===============
\begin{table}
\caption{Characteristics of the best models, with $Y_{\rm i}=0.261$, 
$\alpha_{\rm B}=1.28$ and $t_{\star}=26$~Myrs. 
$T_{\rm c}$ and $\rho_{\rm
c}$ are the central densities and temperatures, given in $10^6$~K and 
g~cm$^{-3}$, respectively. $R_{\rm co}$ is the radius of the convective core and
$R_{\rm ce}$ the radius of the base of the convective envelope. The same
quantities are also given for a calibrated solar model using the same 
input physics
($L_{\sun}=3.846 {\times}
10^{33}$~ergs~s$^{-1}$ and $R_{\sun}=6.9599{\times}10^{10}$~cm).}
\label{tab:int}
\centering
\begin{tabular}{lccc}
\hline
\noalign{\smallskip}
 & EK~Cep~A & EK~Cep~B & Sun \cr
\noalign{\smallskip}
\hline
\noalign{\smallskip}
$T_{\rm c}$ & 21.11 & 12.81 & 15.45 \cr
$\rho_{\rm c}$ & 73.96 & 77.13 & 149 \cr
$T_{\rm eff}$ & 8974 K & 5550 K & 5777 K \cr
$L_{\star}/L_{\sun}$ & 15.74 & 1.465 & 1.000 \cr
$R_{\star}/R_{\sun}$ & 1.644 & 1.311 & 1.000 \cr
$R_{\rm co}/R_{\star}$ & 0.120 & 0.0536 & ... \cr
$R_{\rm ce}/R_{\star}$ & 0.993 & 0.751 & 0.729 \cr
\noalign{\smallskip}
\hline
\end{tabular}
\end{table}
%===============

\begin{figure}[ht]
   \centering
   \includegraphics[height=\figsize,angle=-90]{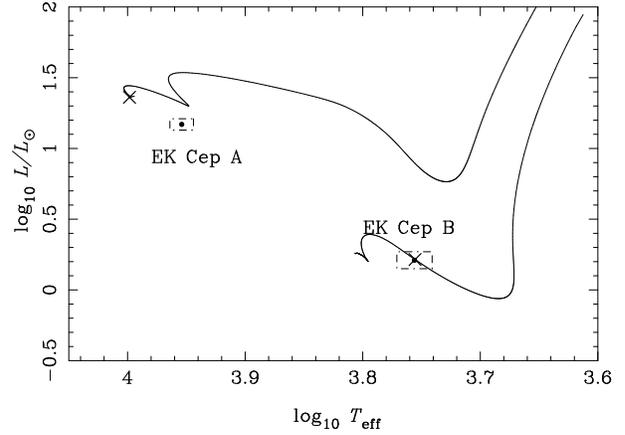}
\caption{Evolutionary tracks of EK~Cep~A and B on the HR diagram. These 
tracks were calculated using $Y_{\rm i}=0.331$ and 
$\alpha_{\rm B}=1.68$, and represent the best fit to the radii of both stars. 
Crosses show the location of 
models with 13.9~Myrs.}\label{fig:hrnew}
\end{figure}

Fig.~\ref{fig:hrnew} shows the evolutionary tracks in the HR diagram of models 
of both components calculated using the parameters given in the previous 
paragraph. 
Since these new parameters fail to improve considerably the fitting 
to the radius of \object{EK~Cep~A}, while the location in the HR diagram 
of the model of 
this star falls well outside the error boxes,
we prefer the solution given in Table~\ref{tab:calib} except for 
$\alpha_{\rm B}$; to fit the radius of \object{EK~Cep~B} 
within the 
observational uncertainties, we choose $\alpha_{\rm B}=1.28$. The age that 
best reproduces the observations is $t_{\star}=26$~Myrs. The characteristics of 
the best models, calculated using these parameters, are shown in 
Table~\ref{tab:int}.

%===============
%===============

The fact that models of \object{EK~Cep~A} depend so 
weakly on the mixing length parameter 
$\alpha$ corresponds to having, in fact, only three ``effective'' 
modelling 
parameters, $Y_{\rm i}$, $\alpha_{\rm B}$ and $t_{\star}$. 
Thus, to fit 
the observables we need an extra parameter, related to the description 
of the interior physics affecting mainly the high mass component. 
This could be the differencial rotation of 
\object{EK~Cep~A}, as described in Y\i ld\i z (\cite{yildiz03}).

%______________________________________________________________________
\section{The role of the time step}\label{sec:time}

Given the importance of \object{EK~Cep~B} in the age determination, 
it is fundamental
that its evolutionary phase (PMS) is adequately calculated.
Consequently, we have performed a detailed analysis on the prescription
that should be adopted for defining the integration time step in the
PMS evolution of this binary.

%===============
\begin{figure}[ht]
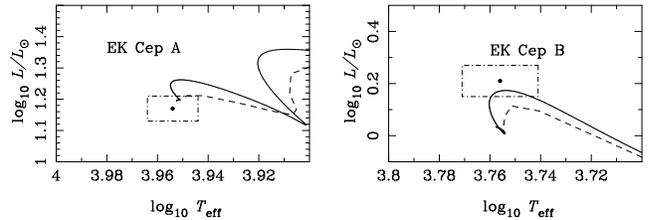

\centering
\begin{tabular}{lr}
\includegraphics[height=\hfigsize,angle=-90]{\figdir/0693fg6a.ps} &
\includegraphics[height=\hfigsize,angle=-90]{\figdir/0693fg6b.ps} \\
\end{tabular}
\caption{The role of the time step. Left panel: EK~Cep~A; in full,
the evolutionary track with a maximum time step of 0.025~Myrs; dashed,
the evolutionary track calculated using the time step prescription appropriate 
for the MS evolution. %without such a strict control (see text for explanation). 
Right panel: the same for EK~Cep~B. In full, the evolutionary track with a 
maximum time step of 0.1~Myrs. }\label{fig:step}
\end{figure}
%===============

Figure~\ref{fig:step} shows again the evolutionary tracks in the HR diagram for
both components of the EK~Cep system.
The effects of an inappropriate choice of the time step are shown;
these effects are only relevant in the PMS phase because in this 
phase the star derives its energy from gravitational contraction; the rate of 
gravitational contraction is incorrect if the time step between two 
consecutive models is too big. The evolutionary tracks represented by a dashed 
line in Fig.~\ref{fig:step} were calculated using a time step prescription that 
is appropriate for the MS evolution: the time step for the next model is 
estimated based on the evolution of the chemical composition. That is, the
faster the evolution of the chemical composition, the shorter the time step 
used. This prescription is not useful for models in the PMS phase, since for 
most of the evolution the chemical composition does not change considerably.
Instead, we must use a prescription that does not allow the
structure of a star to change too much between consecutive time steps.

We can gauge how much the star has contracted
between two consecutive time steps by calculating the
gravitational energy generated between them.
To do this we must
first calculate the model corresponding to $t{+}\Delta t$.
If the gravitational energy generated between the two time
steps ($E_g {\simeq} {-}T \Delta s$, where $s$
is the specific entropy and $T$ is the temperature) is bigger than a given
maximum value, a new model must be calculated corresponding to $t{+}\Delta t$
for a smaller $\Delta t$.
There are, however, some numerical difficulties in this approach if the maximum
gravitational energy allowed to be generated between two consecutive time steps
is too small.

We have chosen a different prescription instead, by using the time
the star spends in the PMS phase ($t_{\rm PMS}$).
This time is expected to scale, roughly, with the
Kelvin-Helmholtz time scale,
%\begin{equation}
$t_{\rm PMS} {\simeq} t_{\rm KH} {\simeq} {G M_{\star}^2} / {2 R_{\star} 
L_{\star}}$,
%\end{equation}
where $R_{\star}$ and $L_{\star}$ are the radius and 
luminosity of the star on the main sequence.
%So, the time a star spends in the PMS goes like,
%\begin{equation}
%t_{\rm PMS} \propto {M_{\star}^2 \over R_{\star} L_{\star}} \;.
%\end{equation}
%Here, $R_{\star}$ and $L_{\star}$ 
These values can be
replaced by their ZAMS values since for most of the time the stars are in the
radiative part of the PMS evolutionary tracks where $R_{\star}$ and $L_{\star}$
do not change very much.

For a given mass range,
$R_{\star}{\propto} M_{\star}^{\xi}$ and
$L_{\star}{\propto} M_{\star}^{\eta}$, on the main sequence.
%, giving that $t_{\rm PMS}{\propto} M_{\star}^{2-\xi -\eta}$.
In the mass range of interest here, $\xi {\simeq} 0.6$ and
$\eta{\simeq} 3.9$ (see Kippenhahn \& Weigert \cite{kippenhahn91}, ch. 22),
it follows that
\begin{equation}
t_{\rm PMS}\propto  {M_{\star}^2 \over R_{\star} L_{\star}}
   \propto  M_{\star}^{2-\xi -\eta}
   \simeq M_{\star}^{-2.5} \;.
\end{equation}
We scale the maximum time step allowed with the time the star
spends in the PMS phase, so that the evolutionary tracks in the PMS phase are
equally divided. For a $1 M_{\sun}$ star, we have considered different values 
for the maximum
time step; for values shorter than about 0.1 Myrs, the evolution of the star
does not change for different values of the time step. Accordingly, the maximum
time step we allow in the evolution during the PMS phase is considered to be 
given by\begin{equation}
\Delta t_{\rm max} = 0.1 \left(\frac {M_{\star}} {M_{\sun}} \right)^{-2.5}
   {\rm Myrs}\,,
\end{equation}
(see also Kippenhahn \& Weigert \cite{kippenhahn91}, ch. 28).

The effects of different prescriptions for the time step
on the calibration of EK~Cep can be seen in Fig.~\ref{fig:step}.
The locus of the calibrated model of EK~Cep~A  (left panel) in the
HR diagram does not change if we do not control the time step
carefully because the star derives most of its energy from nuclear reactions.
On the other hand, the locus of the calibrated model of EK~Cep~B changes
considerably; the luminosity of the model calculated without the restriction on 
the time step is appreciably lower. In fact, we can not fit the observations 
within their error boxes for both stars if we do not restrict the time step in
this way; the luminosity (and radius) of the low mass component is too low.
This partially
explains the results cited in Mart\'{\i}n \& Rebolo (\cite{martin93}),
that the luminosity of EK~Cep~B 
was slightly higher than predicted by pre-main-sequence models. The radius of
our model of EK~Cep~B (with a large time step) is below the
error box in radius, a result also obtained by Claret \etal\  (\cite{claret95}).

The evolution of the models is faster if we do not restrict the
time step. Consequently, instead of the value we found (26.8~Myrs)
the calibrated models of EK~Cep would just have an age of $\simeq 23$~Myrs
when using the same prescription as for the MS evolution.

%______________________________________________________________________
\section{The effect of overshooting}

We also calculated evolutionary models including the effects of convective
overshooting. Convective overshooting is due to the inertia of the convective
elements; since they have a non-zero velocity when they reach the border of the
convective zone, as given by the instability criteria, they overshoot it.
When they reach a radiative zone, they
become colder than the surrounding material (since in a radiative zone
$\nabla < \nabla_{\rm ad}$) and are therefore heated by their
surroundings: they transport energy downwards. To transport all the energy
upwards, the energy transported by the radiation must be greater than the total
luminosity, that is, $\nabla > \nabla_{\rm rad}$. The effect of the 
overshooting is, then, to fix the temperature gradient equal to the adiabatic
gradient and to extend the mixed zone by a certain amount. We used the following
prescription for the extent $d_{\rm ov}$ of the overshooting in the core,
%as been given by $\alpha_{\rm ov} H_{\rm P}$, where 
%for a central convective zone, it is given by 
\begin{equation}
d_{\rm ov} = \alpha_{\rm ov} \times 
   {\rm Min} \left(H_{\rm P},r_{\rm co} \right ) \;,
\end{equation}
where $\alpha_{\rm ov}$ is a free parameter
and $r_{\rm co}$ is the radius of the convective core.

%===============
\begin{figure}[ht]
   \centering
   \includegraphics[height=\figsize,angle=-90]{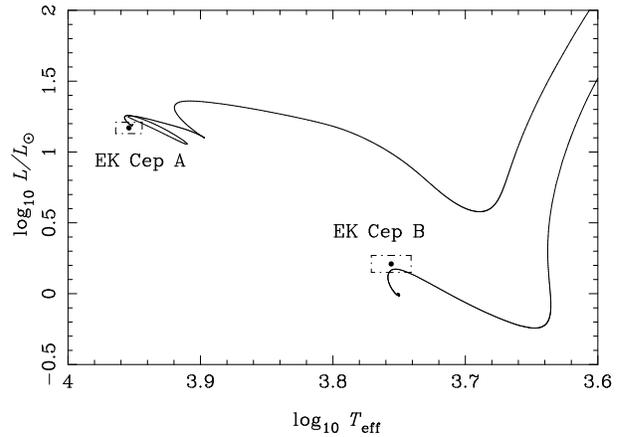}
\caption{Evolutionary tracks of EK~Cep~A and B on the HR diagram. These tracks
were calculated including the effects of convective overshooting, with
$\alpha_{\rm ov}=0.20$.}\label{fig:hrov}
\end{figure}
%===============

Fig. \ref{fig:hrov} shows evolutionary tracks calculated using $\alpha_{\rm
ov}=0.20$. The effects of the overshooting on the tracks are small for the track
corresponding to EK~Cep~B; it merely extends the PMS phase, 
and makes the
drop in luminosity near the end of the PMS phase more abrupt. This is 
caused by the fact that the central convective zone that appears at an age of
about 25~Myrs (see again Fig.~\ref{fig:zc}) has a higher amount of
\element[][12]{C} to burn (the mixed zone extended by overshooting has a higher
mass); the energy produced by \element[][12]{C}-burning is higher, which causes
the luminosity to drop more abruptly, and the fact that there is
more \element[][12]{C} to burn means that the ZAMS will arrive later (by about 
10~Myrs; the precise value depends on the parameter $\alpha_{\rm ov}$).

On the other hand, the track corresponding to EK~Cep~A is considerably changed
with the inclusion of overshooting; it displays an extra ``loop'' after the
first drop in luminosity (compare Figs. \ref{fig:hr} and \ref{fig:hrov}).
This ``loop'' in the evolutionary track in the HR diagram is also seen (but
not discussed) in the evolutionary tracks of Siess \etal\ (\cite{siess00}). It
is caused by the following: as the \element[][12]{C} in the central convective
zone is been burned, its abundance drops and so does the rate of burning. As can
be seen in Fig.~\ref{fig:zc}, the central convective zone recedes towards the
centre, leaving behind a zone of partially depleted \element[][12]{C}. When the
burning of \element[][14]{N} becomes more efficient due to the rising central
temperature, the convective zone expands again, englobing in the mixed central
zone the unburned \element[][12]{C} at the external regions. Now,  
as will be discussed in a subsequent paper
(Marques \etal\, in preparation),
the extent of the
mixed zone due to overshooting leaves more \element[][12]{C} unburned in
these regions; when this second expansion of the convective zone engulfs
fresh \element[][12]{C}, it does so at a rate faster than the burning time.
The rising of the abundance of \element[][12]{C} in the central
convective region (which has now a much higher temperature than it had in the
previous \element[][12]{C}-burning phase) causes a large increase in the
energy production through nuclear reactions in the core, which in turn makes the
central convective region increase very rapidly, with the consequent rapid drop
in the luminosity of the star. This extra ``loop'' in the evolutionary track in
the HR diagram takes some 1.5~Myrs; thus, the arrival on the ZAMS is delayed by
a significant amount (for $\alpha_{\rm ov}=0.2$, about 2.2~Myrs).

For the present work, all these effects are of little importance. The age of the
system
EK~Cep is greater than the age at which these effects take place for a star with
the mass of EK~Cep~A, while it is lower than that at which overshooting
starts to make a (small) difference for stars like EK~Cep~B. Our calibration is
not, therefore, changed by the inclusion of convective overshooting.
An alternative for studying the presence of overshooting in this system is the 
use of asteroseismology (e.g. Monteiro \etal\ \cite{monteiro00}).

%______________________________________________________________________
\section{Conclusions}

EK~Cep is an excellent candidate to test stellar evolutionary models because
it has one member in the MS phase and the other in the PMS phase.
This difference in the evolutionary phase of the components results
in a better calibration of the binary age of 26.8~Myrs.

We have shown that an incorrect treatment of the time step for the evolution
yields models with a lower luminosity during the PMS phase.
If both stars are in the PMS phase, a large time step can be
compensated for by a higher initial helium
abundance (to increase the luminosity of the models;
see Fig.~\ref{fig:var}).
This cannot be done if the stars are in different evolutionary
phases because the increase in the initial helium abundance will make both
stars brighter, while only the PMS component has a lower luminosity, due to the
larger time step.
The dependence of the calibration on this key aspect of the modelling
might explain the results obtained by Claret \etal\  (\cite{claret95}) and other
results cited in Mart\'{\i}n \& Rebolo  (\cite{martin93}), in particular 
that the radius (and luminosity) of EK~Cep~B are
underestimated by evolutionary models. However, we cannot reproduce the 
radius of EK~Cep~A with our models.
The presence of a fast rotating core (as in Y\i ld\i z \cite{yildiz03})
would improve the situation. The use of a correct time step 
solves only part of the problem, by increasing the radius of EK~Cep~B and 
reducing the need for a larger radius of EK~Cep~A.
We stress that our models were calculated using the metallicities obtained by 
Mart\'{\i}n \& Rebolo (\cite{martin93}); if a new determination of the 
metallicity is made, it 
would be possible to discern more clearly the origin of this inconsistency.

In the light of the new calibration, it becomes clear that EK~Cep is
an excellent binary to test different aspects of the physical processes of the 
evolution.
Given the particular configuration of this system, one component
(the primary) is insensitive to the mixing length parameter and age,
while providing a very precise constraint on the helium abundance for
the system.
The effect of the overshooting on
the evolutionary tracks of EK~Cep~A, although not relevant for this work, is
important near the end of the
PMS phase. We will explore the effects of the overshooting during the PMS 
evolution for various types of stars in a subsequent paper 
(Marques \etal\, in preparation).
EK~Cep~B, still in the PMS, provides a very accurate
indication of the age of the system, while being 
independent of the overshooting prescription.
Such a combination makes EK~Cep a test case of how the
detailed observation of young binaries
is of great relevance for improving
the modelling of the evolution in this phase.

%______________________________________________________________________
\begin{acknowledgements}
This work was supported in part by the
{\it Funda\c c\~ao para a Ci\^encia e a Tecnologia}
through project POCTI/FNU/43658/2001. JPM was
supported by grant SFRH/BD/9228/2002 from {\it Funda\c c\~ao para a Ci\^encia e
Tecnologia}.

This work has been performed using the computing facilities provided by the
CESAM stellar evolution code available at

{\tt http://www.obs-nice.fr/cesam/}

We thank the referee, F. Palla, for the many suggestions that improved so 
much the article.
\end{acknowledgements}

%______________________________________________________________________

%______________________________________________________________________
\end{document}